\documentclass[twocolumn,trackchanges]{aastex701}
\usepackage{booktabs}
\usepackage{multirow}

\begin{document}

\title{Dust Seeding Molecules in a Massive Protostar -- Detection of TiO in Orion Source I}

\author{Ziwei E. Zhang}
\email{ziwei.zhang@riken.jp}
\affiliation{Star and Planet Formation Laboratory, RIKEN Pioneering Research Institute, Wako, Saitama 351-0198, Japan}

\author{Nami Sakai}
\email{nami.sakai@riken.jp}
\affiliation{Star and Planet Formation Laboratory, RIKEN Pioneering Research Institute, Wako, Saitama 351-0198, Japan}

\author{Marcelino Ag{\'u}ndez}
\email{marcelino.agundez@csic.es}
\affiliation{Instituto de F\'{\i}sica Fundamental, CSIC, Calle Serrano 123, E-28006 Madrid, Spain}

\author{Linda Podio}
\email{linda.podio@inaf.it}
\affiliation{INAF, Osservatorio Astrofisico di Arcetri, Largo E. Fermi 5, I-50125 Firenze, Italy}

\author{Claudio Codella}
\email{claudio.codella@inaf.it}
\affiliation{INAF, Osservatorio Astrofisico di Arcetri, Largo E. Fermi 5, I-50125 Firenze, Italy}

\author{Tomoya Hirota}
\email{tomoya.hirota@nao.ac.jp}
\affiliation{Department of Astronomical Science, The Graduate University for Advanced Studies, SOKENDAI, 2-21-1 Osawa, Mitaka, Tokyo 181-8588, Japan}
\affiliation{Mizusawa VLBI Observatory, National Astronomical Observatory of Japan, 2-12 Hoshiga-oka, Mizusawa, Oshu-shi, Iwate 023-0861, Japan}

\author{Adam Ginsburg}
\email{adamginsburg@ufl.edu}
\affiliation{Department of Astronomy, University of Florida, PO Box 112055, Gainesville, FL 32611, USA}

\author{Brett McGuire}
\email{brettmc@mit.edu}
\affiliation{Department of Chemistry, Massachusetts Institute of Technology, Cambridge, MA 02139, USA}
\affiliation{National Radio Astronomy Observatory, Charlottesville, VA 22903, USA}

\author{Nadia M. Murillo}
\email{nmurillo@gmail.com}
\affiliation{Instituto de Astronomía, Universidad Nacional Autónoma de México, AP106, Ensenada CP 22830, B. C., México}

\author{Yuki Okoda}
\email{yuki.okoda.05@gmail.com}
\affiliation{ NRC Herzberg Astronomy and Astrophysics, 5071 West Saanich Road, Victoria, BC, V9E 2E7, Canada}

\author{Yoko Oya}
\email{yoko.oya@yukawa.kyoto-u.ac.jp}
\affiliation{ Center for Gravitational Physics, Yukawa Institute for Theoretical Physics, Kyoto University, Kitashirakawa Oiwakecho, Sakyo-ku, Kyoto 606-8502, Japan}

\author{Giovanni Sabatini}
\email{giovanni.sabatini@inaf.it}
\affiliation{INAF, Osservatorio Astrofisico di Arcetri, Largo E. Fermi 5, I-50125 Firenze, Italy}

\author{Aki Takigawa}
\email{takigawa@eps.s.u-tokyo.ac.jp}
\affiliation{Department of Earth and Planetary Science, The University of Tokyo, 7-3-1 Hongo, Tokyo 113-0033, Japan}

\author{Kei E. I. Tanaka}
\email{kt503i@gmail.com}
\affiliation{Department of Earth and Planetary Sciences, Institute of Science Tokyo, Meguro, Tokyo, 152-8551, Japan}

\author{Yichen Zhang}
\email{yczhang.astro@gmail.com}
\affiliation{Department of Astronomy, School of Physics and Astronomy, Shanghai Jiao Tong University, 800 Dongchuan Road, Shanghai 200240, China}
\affiliation{State Key Laboratory of Dark Matter Physics, School of Physics and Astronomy, Shanghai Jiao Tong University, Shanghai 200240, China}
\affiliation{Key Laboratory for Particle Astrophysics and Cosmology (MOE) / Shanghai Key Laboratory for Particle Physics and Cosmology, Shanghai 200240, China}

\author{Qiuyi Luo}
\email{nami.sakai@riken.jp}
\affiliation{Institute of Astronomy, Graduate School of Science, The University of Tokyo, 2-21-1 Osawa, Mitaka, Tokyo 181-0015, Japan}
\affiliation{Department of Astronomy, School of Science, The University of Tokyo, 7-3-1 Hongo, Bunkyo, Tokyo 113-0033, Japan}

\author{Yu Cheng}
\email{ycheng.astro@gmail.com}
\affiliation{National Astronomical Observatory of Japan, 2-21-1 Osawa, Mitaka, Tokyo 181-8588, Japan}

\author{Izaskun Jim{\'e}nez-Serra}
\email{ijimenez@cab.inta-csic.es}
\affiliation{Centro de Astrobiología (CAB, CSIC-INTA), Carretera de Ajalvir km 4, Torrejón de Ardoz, E-28850 Madrid, Spain}

\author{Shogo Tachibana}
\email{shogotachibana@g.ecc.u-tokyo.ac.jp}
\affiliation{Department of Earth and Planetary Science, The University of Tokyo, Hongo, Tokyo 113-0033, Japan}

\begin{abstract}
We report the first detection of TiO in star-forming regions based on Atacama Large Millimeter/submillimeter Array observations of Orion Source I, a well-characterized massive protostar. Multiple rotational transitions are identified, with emission spatially resolved within $\sim 50$ au, showing a compact distribution with a velocity structure consistent with the base of a rotating outflow. The spatial and velocity distributions of TiO are consistent with those of AlO, with both species being key dust seeding refractory molecules. The column density of TiO is derived to be $(3.0 \pm 0.4)\times10^{15}\ {\rm cm^{-2}}$, corresponding to $X_{\rm TiO/SiO} \sim 12.8 \pm 1.7 \times10^{-3}$, higher than CI chondrites $X_{\rm Ti/Si}$ value and indicative of efficient dust-to-gas conversion near the protostar. 
The detection of TiO, a key dust seeding species, offers important constraints on refractory chemistry and the formation environments of primitive minerals, linking astrochemical processes in protostellar systems to the earliest stages of Solar System material formation.
\end{abstract}


\section{Introduction} 
\label{sec: Intro}

Refractory and salt species have been identified to be associated with the highly-heated or -shocked material in star-forming regions \citep{Plambeck2016, Hirota2017, Ginsburg2019salt, Tachibana2019, Tanaka2020, Ginsburg2023, Law2023}. In particular, highly refractory molecules, such as SiO and AlO, have been observed in Orion Source I (SrcI hereafter) and suggested to be good probes of the innermost region and/or the outflow launching point of massive protostars \citep{Hirota2017,Tachibana2019,Kim2019}. 

Despite refractory molecules serving as seeds for interstellar dust and protoplanetary solids, constraining the chemical composition of dust in protostellar systems remains challenging. In contrast, primitive Solar System materials provide direct evidence for refractory solids formed under high-temperature conditions \citep{Lodders2003}. In planetary science, minerals enriched with refractory elements (e.g., Ca, Al, and Ti) are commonly referred to as Ca- and Al-rich inclusions (CAIs). These are considered the oldest solids in the Solar System and formed under high-temperature \citep{Connelly2012}. However, it is still under debate whether they originated locally in the disk of protostars. This is partly due to the limited observations on refractory dust seeding molecules in star-forming regions. 


SrcI ($\sim8.7$ M$_\odot$, \citealt{Hirota2017}; $\sim15$ M$_\odot$, \citealt{Ginsburg2018}), located in the Orion KL region ($\sim415$ pc, \citealt{Menten2007,Kounkel2018}), is one of the best-characterized massive protostars at disk-scale. It is associated with prominent SiO and H$_2$O masers \citep{Goddi2009, Hirota2018, Niederhofer2012, Issaoun2017} and drives a rotating outflow which is probed by Si$^{18}$O \citep{Hirota2017}. \cite{Ginsburg2019salt} reported rich detection of NaCl and KCl, as well as their isotopologues such as $^{41}$KCl, in its disk. Moreover, AlO, a highly refractory molecule which is considered as a key dust forming molecule, is detected at the outflow launching point of SrcI \citep{Tachibana2019}. Such detections were only limited to Asymptotic Giant Branch (AGB) stars previously and motivate further studies of dust sublimation/formation with highly refractory molecules (e.g., TiO) in protostellar environments \citep{Tenenbaum2009,Decin2017,Takigawa2017}. 

TiO is observed in oxygen-rich AGB stars (e.g., \citealt{Kaminski2013,Ohnaka2025}) and is considered as one of the dust-seeding molecules, as it acts as the precursor of TiO$_2$, which plays a key role in dust nucleation at temperatures $\gtrsim 1000$ K \citep{Gail1998}. With the high binding energy (672.4 kJ mol $^{-1}$), TiO is the dominant Ti-bearing species in the highly-heated environments, 
making it an ideal proxy for elemental abundance estimation under such conditions \citep{Gail1998}. AlO is commonly detected together with AlOH in AGB stars, as they are chemically coupled \citep{Mangan2021}. Both molecules are abundant in oxygen-rich circumstellar gas and considered to be gas-phase precursors of stoichiometric (Al$_2$O$_3$)$_n$ clusters \citep{Mangan2021, Gobrecht2022}. 
Therefore, a comprehensive understanding of refractory chemistry and dust formation requires characterization of both Ti- and Al-bearing species, which serve as complementary indicators of dust forming pathways.

In this letter, we report the first detection of TiO in star-forming regions based on the recent Atacama Large Millimeter/submillimeter Array (ALMA) observations of SrcI. The emission is spatially resolved and concentrated at the outflow base. In addition, one possible AlOH transition is identified in Appendix \ref{sec: AlOH}. These results provide new constraints on refractory dust processing near protostars and the formation environments of CAI-like materials.

\section{Observation}
\label{sec: Obs}

\begin{deluxetable*}{lccccccccc}
\tabletypesize{\scriptsize}
\label{tab: linelist}
\tablecaption{Line List $^a$ \label{tab:linelist}}
\tablehead{
\colhead{Mol.} &
\colhead{Freq.} &
\colhead{Trans.} &
\colhead{log($A$)} &
\colhead{$E_{\rm u}$} &
\colhead{$g_{\rm u}$} &
\colhead{$\theta_{\rm Beam}$} &
\colhead{$V_{\rm peak}$/$\Delta V_{\rm FWHM}$ $^{b}$} &
\colhead{Intensity $^{c}$} \\
\colhead{}&
\colhead{(MHz)} &
\colhead{} &
\colhead{(s$^{-1}$)} &
\colhead{(K)} &
\colhead{} &
\colhead{($\!\!^{\prime\prime}$)} &
\colhead{(km s$^{-1}$)} &
\colhead{(Jy beam$^{-1}$ km s$^{-1}$)}
}
\startdata
\multirow{8}{*}{TiO} & 348159.79 (0.04) & $ J=11-10, \Omega=1$ & $-$2.59 &
98.74 & 23 & $ 0.38 \times 0.29$ & $8.8\pm0.3/12.8\pm0.7$ & $2.75 \pm 0.10$ \\
\cline{2-9}
 & 664340.12 (0.03) & $J=21-20, \Omega=1$ & $-$1.73 & 349.31 & 43 & $0.20 \times 0.19$ & 
$8.1\pm0.2/13.0\pm0.6$ & $2.66\pm0.21$ \\
\cline{2-9}
 & 853761.70 (0.06) & $J=27-26, \Omega=1$ & $-$1.40 & 572.44 & 55 & $0.09 \times 0.09$ & 
$7.6\pm0.1/11.7\pm0.3$ & $2.86\pm0.05$ \\
\cline{2-9}
 & 355623.29 (0.01) & $J = 11-10, \Omega =3 $ & $-$2.59 & 386.19 & 46 & $0.33 \times 0.28$ & 
\textit{8.0/13.0} & Blend (KCl, others) \\
\cline{2-9}
 & 484796.02 (0.01) & $J = 15-14, \Omega=3$ & $-$2.16 & 469.96 & 62 & $ 0.15 \times 0.11$ & \textit{8.0/13.0} & Blend (iCOMs) \\
\cline{2-9}
 & 646071.27 (0.03) & $J = 20-19,\Omega =3 $ &
$-$1.78 & 609.52 & 82 &
$ 0.24 \times 0.18$ &
\textit{8.0/13.0} & Blend (SiS, SO$_2$) \\
\cline{2-9}
 & 189929.83 (0.00) & $ J = 6-5, \Omega =1$ & $-$3.40 & 30.38 & 13 & $0.13 \times 0.10$ & ...& iCOMs Emission \\
\cline{2-9}
 & 323313.48 (0.04) & $J = 10- 9, \Omega = 3$ &
$-$2.72 & 369.13 & 42 & $0.23 \times 0.16$ & ...& iCOMs Emission \\
\cline{2-9}
 & 474687.81 (0.01) & $ J = 15-14, \Omega =1$ & $-$2.17 & 180.76 & 31 & $ 0.17 \times 0.16$ & ...& H$_2$O absorption \\
\tableline
AlOH $^d$ & 660175.91 (0.43) & $J = 21-20$, FS & $-$2.75
& 348.75 & 43 & $0.22 \times 0.15$ &$8.5\pm0.2/12.0\pm0.6$ & $1.37\pm0.70$ \\
\tableline
\enddata
\tablerefs{
$^{a}$ Line information is taken from the Cologne Database of Molecular Spectroscopy (CDMS; \citealt{CDMS16}). The transitions at $189929.83$ MHz, $474687.81$ MHz and $323313.48$ MHz are heavily blended with other strong emission or absorption features (e.g., iCOMs and H$_2$O) and overall no distinguishable emission attributable to TiO is detected above $3\sigma$. The $\Lambda$ doubling is unresolved in our observations, and the corresponding pairs are treated as single transitions in the analysis. $^b$ Parameters in $italic$ are to demonstrate the emission features, not the true linewidths. $^c$ Mean flux within a 100 au region centered at the continuum peak. $^d$ Possible AlOH transition, the fitting baseline is tuned as $\pm 25$ km s$^{-1}$ due to the strong emission at nearby frequencies.}
\end{deluxetable*}

We analyze all archival ALMA datasets available to date that cover TiO transitions and have comparable angular resolutions ($\sim0.^{\prime\prime}1$--$0.^{\prime\prime}4$), including projects 2012.1.00123.S, 2013.1.00048.S, 2016.1.00970.S, 2023.1.01382.S, 2025.1.00274.S, and 2025.1.00639.S. The data is processed with the Common Astronomy Software Applications (CASA, \citealt{CASA2022}). Continuum channels are determined with STATCONT from dirty images for each dataset \citep{Sanchez-Monge2017}. Self-calibration for both phase and amplitude is applied to obtain better data quality. Both the continuum and spectral images are prepared with Briggs robustness of 0.5. To focus on the disk region and resolve out the extended components, we only use the visibility data with UV distance longer than 400 $k\lambda$ for continuum imaging \citep{Hirota2016}. The Band 10 data were processed separately. Continuum channels were selected manually, and the imaging are with Briggs robustness of 2.0 \citep{Hirota2026}.
Continuum for Band 7 is kept with the original visibility data due to the relatively lower spatial resolution. The resulting synthesized beam sizes are listed in Table \ref{tab: linelist}. The phase centers of the observations are $(05^{\mathrm h}35^{\mathrm m}14^{\mathrm s}.514,\, -05^\circ22'30.''556)$ for 2012.1.00123.S; $(05^{\mathrm h}35^{\mathrm m}14^{\mathrm s}512,\, -05^\circ22'30.''570)$ for both 2013.1.00048.S and 2016.1.00970.S; $(05^{\mathrm h}35^{\mathrm m}14^{\mathrm s}517,\, -05^\circ22'30.''613)$ for 2023.1.01382.S; and $(05^{\mathrm h}35^{\mathrm m}14^{\mathrm s}518,\, -05^\circ22'30.''619)$ for both 2025.1.00274.S and 2025.1.00639.S. For simplicity, spatial distributions are presented as offsets in au relative to the continuum peak position.

\section{Results and Discussion}
\label{sec: Results}

The molecular identifications presented in this work rely on spectral, spatial, and kinematic information. Each candidate transition is required to satisfy multiple observational constraints: (1) appearing at the expected rest frequency, (2) originating from the same compact region, (3) exhibiting a consistent velocity gradient, and (4) showing spatial and kinematic consistency with chemically related refractory species (e.g., AlO, which traces the outflow base; \citealt{Tachibana2019}). Accordingly, spectral coincidence, spatial morphology, and position–velocity (PV) diagrams are considered jointly when evaluating molecular identifications, with the latter providing independent constraints beyond the spectra alone. This approach is particularly well suited to spatially resolved observations of well-characterized sources such as SrcI, where the expected morphology and kinematics substantially reduce the ambiguity introduced by spectral blending.

\subsection{Detection of TiO}
\label{subsec: TiO}

\begin{figure*}[htbp]
\centering
\includegraphics[width=1\textwidth, trim= 0 1cm 16cm 0, clip]{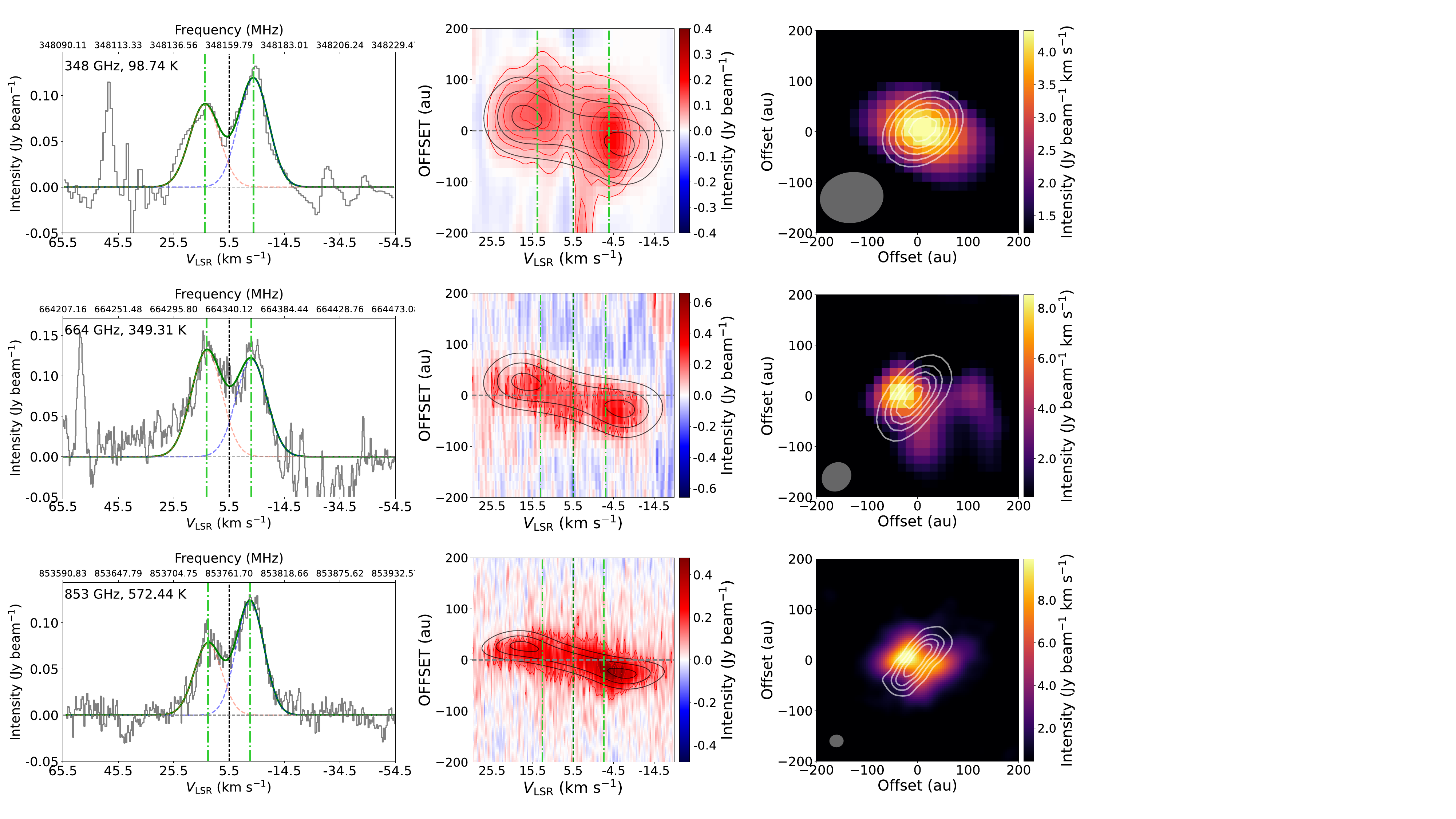}
\caption{Left: Mean spectra of detected TiO transitions within 100 au from the continuum center. These mean spectra include both velocity peaks seen in the PV diagrams. The Gaussian fits of the two velocity components and the overall profile are indicated with the dashed and solid lines, respectively. The peak velocity offsets ($V_{\rm peak}$) are marked with light green dash-dot lines. $V_{\rm peak}$ is the velocity of the measured symmetric peak offset, i.e., the two peaks are at $\pm V_{\rm peak}$ km s$^{-1}$ from the system velocity (5.5 km s$^{-1}$). The Gaussians are fit with the constraints that both components are with the same $V_{\rm peak}$. The linewidths ($\Delta V_{\rm FWHM}$) and peak offsets ($V_{\rm peak}$) are reported in Table \ref{tab: linelist}. 
Middle: PV diagrams overlaid with a Keplerian rotation model at at at 30\%, 60\%, 90\% of the peak intensity (see Section \ref{subsec: TiO}). The red contours are plotted at intervals of $3\sigma$, starting from $3\sigma$, for the top and bottom panels. For the top panel, the contours are plotted at intervals of $15\sigma$, starting from $15\sigma$. The feature with low velocity gradient at $\sim -200$ au for TiO at 348 GHz is from the extended material. Right: Spatial distribution. At 415 pc, 100 au = $0.^{\prime\prime}24$.
From top to bottom, the emission is masked below $100\sigma$, $5\sigma$, and $3\sigma$; while the continuum is shown in white contours starting at $300\sigma$, $10\sigma$,  and $10\sigma$. The RMS of the molecular emission are $2.0\times10^{-3}$, $5.5\times10^{-2}$,  and $4.0\times10^{-2}$ Jy beam $^{-1}$, while those of the continuum are $1.0\times10^{-3}$, $1.1\times10^{-2}$,  and $5.0\times10^{-3}$ Jy beam $^{-1}$.
}
\label{fig: TiO1}
\end{figure*}

All TiO transitions within the wavelength ranges of the analyzed datasets are listed in Table \ref{tab: linelist}. Among the nine rotational transitions, three are unblended, three are partially blended, and three cannot be identified due to severe blending and/or insufficient sensitivity. In Fig. \ref{fig: TiO1}, we show the mean spectra within 100 au from the continuum peak of the three unblended transitions. We fit the lines with a two-component Gaussian in which the offset from system velocity (5.5 km s$^{-1}$) is symmetric, i.e., there are two components at $\pm V_{\rm peak}$ velocity offset, with equal linewidths, but the amplitude is allowed to vary independently.
The PV diagrams and spatial distributions are also shown. TiO emission is compact within $\sim 50-100$ au near the central object with a ``cone-shape" enhancement towards the northeast seen with the 664 GHz and 853 GHz transitions, likely due to the slight inclination of the disk. 
The localized ``cone-shape" and large velocity gradient ($\sim -20$ km s$^{-1}$ -- $+20$ km s$^{-1}$) suggest that TiO is present at the outflow base or launching point. Such emission features are also consistent with the previous detection of AlO, another key refractory molecule, in SrcI \citep{Tachibana2019}. The rotating outflow of SrcI has been reported in \cite{Hirota2017} and therefore we fit the rotation structure in the PV diagrams with a simple Keplerian model \citep{Oya2022}. The PV diagrams are taken with a 0.$^{\prime\prime}$20 cut along the disk mid-plane (PA = $36^{\circ}$, \citealt{Bally2011, Ginsburg2018}). The low velocity gradient feature ($\lesssim 3-5$ km s$^{-1}$) seen at $\sim -200$ au in the PV diagram of the 348 GHz transition is from extended material. The overlaid model contours have been convolved with the corresponding observational beams. Considering the emission is at the outflow base, the model is used as an approximate description and the derived mass should be interpreted as an effective mass inferred from the angular momentum of the outflowing material. As a result, we estimate a disk (or a flat ring) with an outer radius $\sim$50 au, an inner radius $\sim$15 au, and a central mass $\sim$10 M$_{\odot}$. The estimated central mass  is between the previous derivation based on H$_2$O emission at $\sim 15$ M$_{\odot}$ and the one derived with Si$^{18}$O at $\sim 8.7$ M$_{\odot}$ \citep{Ginsburg2018, Hirota2017}. This  further suggests that TiO is tracing the outflow base, as it has been suggested that SiO is tracing the outflow and H$_2$O is tracing the disk \citep{Hirota2017}. The 1 GHz spectra around the TiO lines and the analysis of the three blended transitions are shown in Appendix as references to further confirm detection. 

\subsection{Implications for Refractory Dust Processing}
\label{subsec: Discussion}

\begin{figure}[htbp]
\centering
\includegraphics[width=0.47\textwidth, trim= 0 0 0 0, clip]{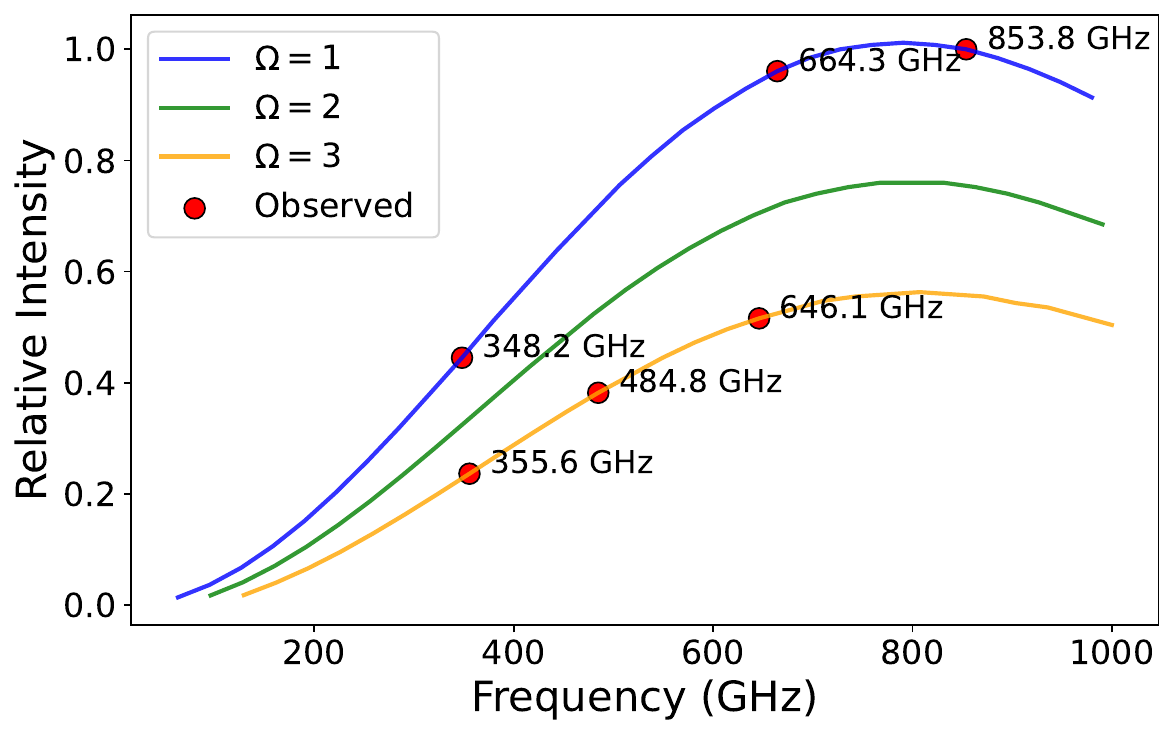}
\caption{Relative TiO line intensities as a function of frequency from LTE calculations at $483.3$ K (Section \ref{subsec: Discussion}). The intensities are normalized to the TiO 853 GHz transition. The transitions presented in this work are marked in red. The $\Omega =3$ transitions, although blended, have intensities $\sim20\% - 50\%$ of that of the 853 GHz transition under LTE conditions.}
\label{fig: LTE}
\end{figure}

Assuming the gas is in local thermodynamic equilibrium (LTE) and optically thin, we estimate the TiO abundance in the central region with the measured intensities using excitation analysis of the three unblended transitions (Table  \ref{tab: linelist}) \citep{Goldsmith1999}. With a derived rotation temperature of $(484.3 \pm 73.0)$ K, we obtain an average TiO column density of $(3.0 \pm 0.4)\times10^{15}\ {\rm cm^{-2}}$.
Here, the rotation temperature is the effective excitation temperature, likely to be different from the kinetic temperature of the gas which is heated by the central protostar. In Fig. \ref{fig: LTE}, we show the normalized TiO line intensities as a function of frequency from standard LTE calculations at 484.3 K, adopting the molecular data from CDMS \citep{CDMS16}. 
The intensities are calculated assuming optically thin emission, identical linewidths, and the same observational resolution and sensitivity. The detected $\Omega=1$ transitions are among the strongest TiO lines, while the $\Omega=3$ transitions have calculated intensity of $\sim20\%-50\%$ of that of the 853 GHz transition. Notably, the $\Omega=3$ transitions, despite being blended, could exhibit relatively higher emission (e.g., at 646.1 GHz, Fig. \ref{fig: TiO_blended}).

Taking the SiO column density to be $2.35\times10^{17}\ {\rm cm^{-2}}$ \citep{Ziurys1987}, we find an abundance ratio of $X_{\rm TiO/SiO} = (12.8 \pm 1.7)\times10^{-3}$. Such a value is higher to the CI chondrites X$_{Ti/Si}$ at $\sim 2.45\times10^{-3}$ \citep{Asplund2021}. The elevated $X_{\rm TiO/SiO}$ ratio could be partially attributed to the high abundance of other Si-bearing molecules, such as SiS, which is also commonly detected in SrcI. Moreover, the reported SiO abundance was derived from observations with a much lower angular resolution ($61^{\prime\prime}$) -- the abundance in the central region could be substantially higher. Therefore, our derived abundance ratio could be considered as an upper limit. Nevertheless, the inferred abundance and ratio imply efficient conversion of Ti- and Si-bearing dust to the gas phase around the disk region. We note that this estimate is approximate, as only three transitions are applied and the physical conditions of the outflow base (high temperature and radiation) likely drive the gas out of LTE. 

Previous studies of SrcI have shown that refractory species (e.g., SiO and SiS) are enhanced in the outflow, while salt species (e.g., NaCl and KCl) are more concentrated in the disk \citep{Hirota2012,Ginsburg2019salt,Wright2024}. Volatile species, such as H$_2$O and SO$_2$, can show either extended emission or be concentrated near the disk, depending on their excitation conditions \citep{Plambeck2016,Wright2024}. \cite{Tachibana2019} have suggested that AlO is emitting from the launching point of the outflow based on its spatial distribution.

TiO shows a spatial and velocity distribution similar to that of AlO. The simple rotation model suggests that TiO kinematics are consistent with Si$^{18}$O, a tracer of the rotating outflow \citep{Hirota2017}, and distinct from high-excitation H$_2$O emission associated with the rotating disk \citep{Ginsburg2018}. The compact emission of AlO and TiO at the outflow base likely reflects their refractory nature and origin in the hottest inner regions. As the gas cools (AlO $\lesssim 1650$ K and TiO $\lesssim 1450$ K at $10^{-5}$ bar), such as further out in the outflow and/or in the disk mid-plane, these species are expected to condense into corundum (Al$_2$O$_3$) and perovskite (CaTiO$_3$) \citep{Ebel2023}. AlO and AlOH are chemically coupled through H$_2$O/H$_2$ reactions \citep{Gobrecht2022}, which could also be efficient in star-forming regions.


In planetary science, CAIs are interpreted as condensates from the solar nebula, but the processes governing the destruction and sublimation of interstellar dust, as well as its eventual re-condensation in the vicinity of young stars, are not yet fully understood. It has also been proposed that the isotopic fractionation and rare earth element signatures in a subset of CAIs result from post-formational evaporation at extreme temperatures \citep{Grossman2008}. Identifying the timing and environment of dust sublimation is therefore critical for a comprehensive understanding of the chemical and physical evolution of early solar system materials.

The detection of dust seeding refractory molecules is crucial for understanding the hot inner region of not only SrcI but also for protoplanetary systems in general. These species help us to characterize the temperature, density, and ionization structure of these highly heated and irradiated innermost zones where dust is sublimated and/or destroyed. Moreover, such studies will further help constrain the condensation/sublimation front of refractory minerals and the formation environments of CAI-like materials in protostellar systems, linking astrochemical signatures to the earliest stages of planetary material formation in the solar system.

\section*{Acknowledgements}
The authors thank the anonymous referees for their constructive comments. This letter makes use of the following ALMA data: ADS/JAO.ALMA\# 2012.1.00123.S (PI: Richard Plambeck ), 2013.1.00048.S (PI: Tomoya Hirota), 2016.1.00970.S (PI: Tomoya Hirota), 2025.1.00274.S (PI: Nami Sakai), and 2025.1.00639.S (PI: Ziwei Zhang). ALMA is a partnership of the ESO (representing its member states), the NSF (USA) and NINS (Japan), together with the NRC (Canada) and the NSC and ASIAA (Taiwan), in cooperation with the Republic of Chile. The Joint ALMA Observatory is operated by the ESO, the AUI/NRAO, and the NAOJ. This study is supported by a Grant-in-Aid from the Ministry of Education, Culture, Sports, Science, and Technology of Japan (26K17208, JP20H05844, JP20H05845 and JP25H00676) and a pioneering project in RIKEN (Evolution of Matter in the Universe and Basic Science in Space Bridged by Cutting-Edge Space Utilization Technology). N.M.M. acknowledges support from the DGAPA–PAPIIT IA103025 grant. CC, LP and GS acknowledge the project PRIN MUR 2022 FOSSILS (Prot. 2022JC2Y93), the project ASI-Astrobiologia 2023 MIGLIORA (F83C23000800005), the INAF-GO 2023 fundings PROTO-SKA (C13C23000770005), and the INAF-Minigrant 2023 TRIESTE (PI: G. Sabatini).
\facilities{ALMA.}
\software{CARTA \citep{Carta}, CASA \citep{CASA2022}, astropy \citep{astropy:2013, astropy:2018, astropy:2022}, spectral-cube\citep{Ginsburg2019}, and pvextractor \citep{Ginsburg2016}.}

\appendix
\section{Spectra}

\begin{figure*}[htbp]
\centering
\includegraphics[width=1\textwidth, trim= 0 6cm 4cm 0, clip]{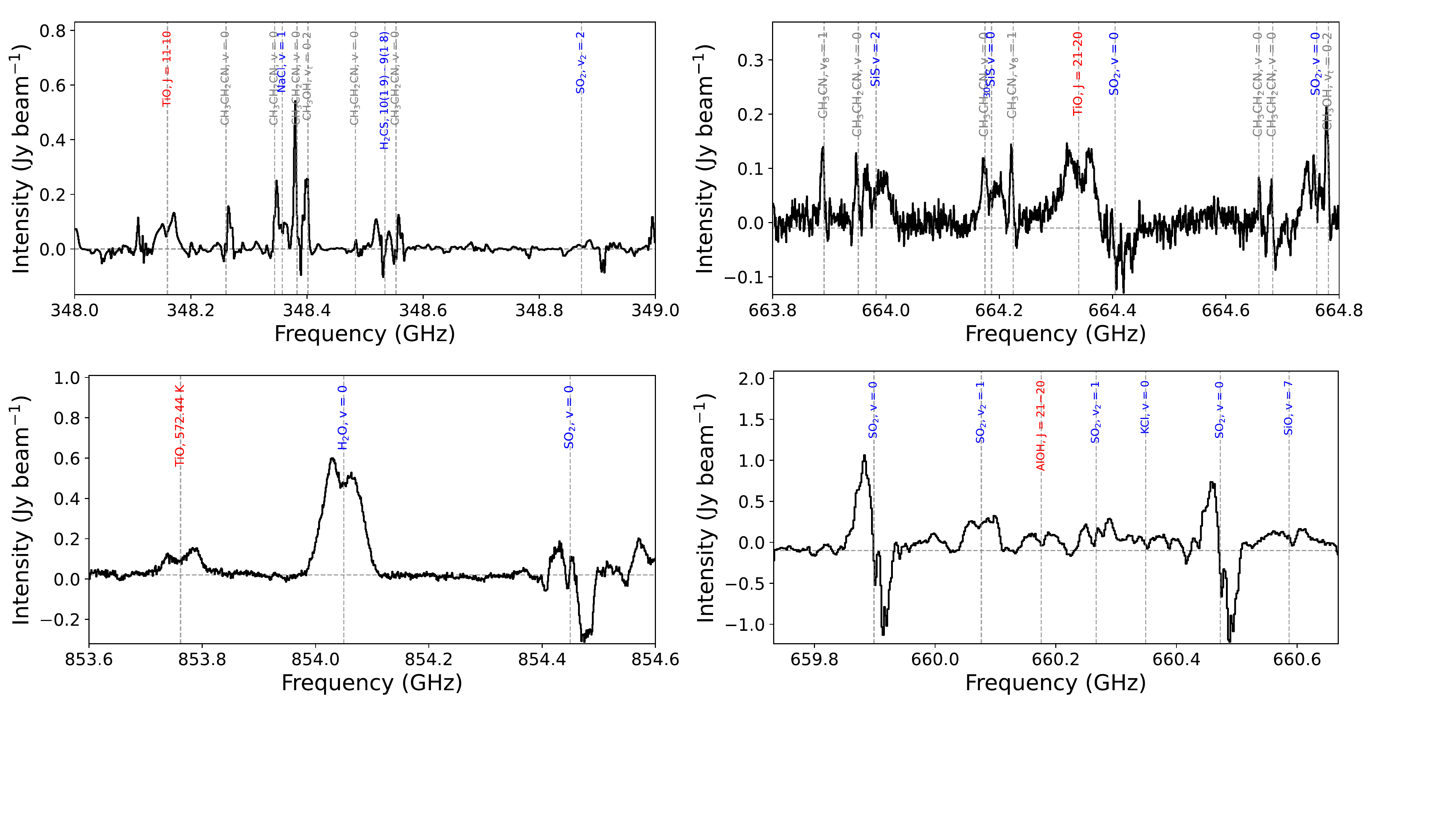}
\caption{The 1 GHz-wide spectra of the detected TiO and possibly observed AlOH lines. The mean spectra are extracted from a circular area within 100 au from the continuum center. The TiO and AlOH transitions are marked in red, iCOM transitions in gray, and other molecular transitions in blue. The baselines are tuned with spectral fittings.}
\label{fig: spectra}
\end{figure*}

Fig. \ref{fig: spectra} presents the 1 GHz-wide mean spectra within 100 au of the detected TiO and possibly observed AlOH lines. These transitions are among the most prominent features in the spectra, with intensities comparable to those of commonly detected species such as H$_2$O, SiO, and SO$_2$. The broader linewidths and double-peak profiles of the TiO emission, comparable to those of disk/outflow tracing molecules such as H$_2$O and SiS, suggest that TiO originates around the disk region. In contrast, some iCOMs (e.g., CH$_3$CN), which arise from the extended material, are identified and exhibit much narrower linewidths ($\lesssim 3-5$ km s$^{-1}$), distinguishable from disk tracers. 

\section{TiO Blended Transitions}
\label{subsec: TiO_blended}

All three TiO lines show consistent emission features and kinematics structures. As the spectra are complex due to line blending and contamination, the TiO transitions are more reliably identified from the PV diagrams, which reveal consistent compact, double-peaked emission and similar velocity gradients. The 484 GHz transition is enhanced at the blue-shifted and blended with extended emission ($\gtrsim 150$ au and very narrow velocity range of $\lesssim 3-5$ km s$^{-1}$) from species such as iCOMs. For the lines that are more blended with the emission from the disk region ($\lesssim 100$ au), we mark the expected velocities of the blending species with dashed lines. The 646 GHz transition is blended with SiS ($E_u = 2708.96$ K) and SO$_2$ ($E_u = 1286.44$ K), both of which are concentrated in the disk region. Clear blue-shifted peaks with high intensities are observed for TiO, SiS, and SO$_2$, while the corresponding red-shifted components are truncated at the edge of the spectral window. The 355 GHz transition is affected by blending from both compact emission (KCl, $E_u = 849.9$ K) and extended emission. Only the blue-shifted components of TiO and KCl can be distinguished, and both peaks are clearly defined. Overall, all three blended TiO transitions exhibit consistent velocity structures and compact emission ($\lesssim 100$ au).

\begin{figure*}[htbp]
\centering
\includegraphics[width=1\textwidth, trim= 0 14cm 0 0, clip]{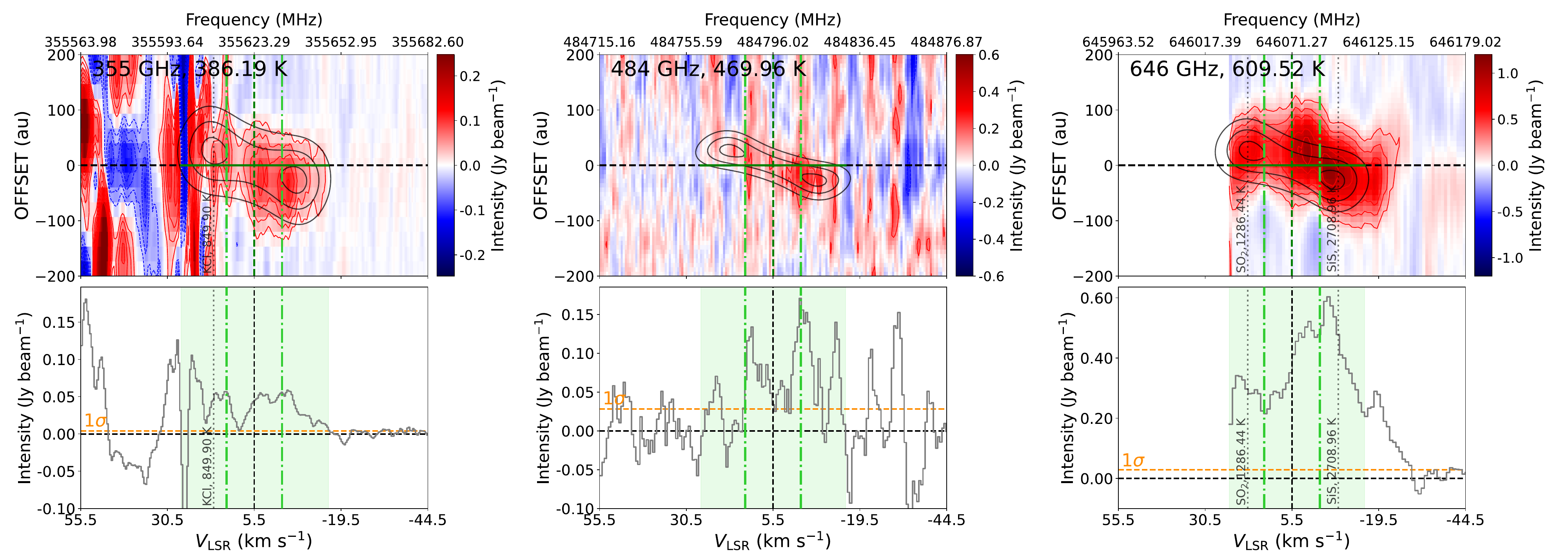}
\caption{PV diagrams and spectra for the three blended TiO transitions. The red contours are plotted at intervals of $4\sigma$ , starting from $4\sigma$.
The mean spectra are extracted from a circular area within 100 au from the continuum center. At 415 pc, 100 au = $0.^{\prime\prime}24$. The effective emission range ($V_{\rm peak} = 8.0$ km s$^{-1}$  and $\Delta V_{\rm FWHM} = 13$ km s$^{-1}$ , Table \ref{tab: linelist}) of TiO is shaded with light green. Noise levels at $1\sigma$ are marked with orange dashed lines. The spectrum at 646 GHz is truncated at the edge of the spectral window. The 484 GHz transition is blended with extended emission, characterized by low-velocity-gradient features at offsets of $\gtrsim \pm 100$ au. The 646 GHz transition is blended with compact emission from SiS and SO$_2$, while the 355 GHz transition is contaminated by both extended and compact emission, including KCl. From left to right, the RMS of the molecular emission are $5.5\times10^{-5}$, $4.0\times10^{-2}$, and $4.0\times10^{-2}$  Jy beam $^{-1}$.}
\label{fig: TiO_blended}
\end{figure*}

\section{Possible AlOH Transition}
\label{sec: AlOH}

\begin{figure*}[htbp]
\centering
\includegraphics[width=1\textwidth, trim= 0 22cm 0cm 0, clip]{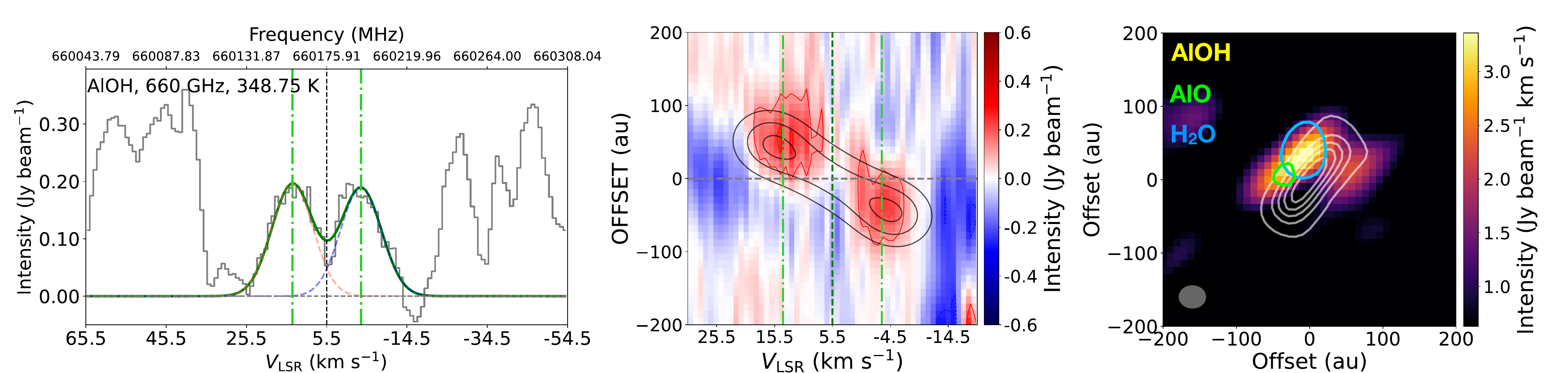}
\caption{Possible AlOH J=21-20 transition (fine structure) at 660 GHz. The spectrum is fitted with the same procedure as TiO (see Fig. \ref{fig: TiO1}), and the fitting parameters are reported in Table \ref{tab: linelist}. The PV diagram is overlaid with a Keplerian rotation model at 30\%, 60\%, and 90\% of the peak intensity (see Section 3.2). Red contours are plotted at intervals of $3\sigma$, starting from $3\sigma$.  At 415 pc, 100 au = $0.^{\prime\prime}24$. The spatial distribution is masked below $3\sigma$ and is overlaid with the emission peaks of AlO and H$_2$O ($\geq 95\%$ of the peak intensity). The continuum in white contours starts at $10\sigma$. The channel RMS of the AlOH emission is $4.0\times10^{-2}$ Jy beam $^{-1}$ and the continuum RMS is $5.0\times10^{-3}$ Jy beam $^{-1}$.}
\label{fig: AlOH}
\end{figure*}

One possible AlOH transition (J = 21 -- 20, 660 GHz, fine structure, $E_u = 348.75$ K) is identified (see Fig. \ref{fig: AlOH} for spectrum, PV diagram, and spatial distribution). The PV diagram shows clear rotation features but a smaller velocity gradient than that of TiO. We overlay a Keplerian model with an outer radius of $\sim$65 au, an inner radius of $\sim$50 au, and a central mass of $\sim$10 M$_{\odot}$, suggesting the emission could be more extended than TiO \citep{Oya2022}. We also compare this line with the emission peaks of AlO (N = 17 -- 16, $E_u = 280.97$ K) and H$_2$O ($v_{2} = 1, 4_{2,2}-3_{3,1}, E_u = 2766.63$ K) \citep{Hirota2017, Hirota2018,Tachibana2019}. The possible AlOH emission is distributed over a larger radius ($\sim100$ au) and is primarily concentrated along the disk surface. The emission peaks of AlO and TiO are shifted to the southeast, whereas that of the possible AlOH emission is shifted to the northwest and is correlated with H$_2$O peak. If the identification is correct, this distribution could provide clues to the formation of AlOH. AlOH may be produced via reactions between AlO and H$_2$ and/or H$_2$O, both of which are abundant around the SrcI disk \citep{Mangan2021,Gobrecht2022}. The extended emission along the disk surface, the spatial offset from the AlO peak, and the similar northwest shift seen in H$_2$O could be explained by this reaction route. Moreover, according to chemical equilibrium models, AlOH ($\gtrsim 1300$ K) favors lower temperatures than AlO ($\gtrsim 1500$ K), which could explain the extended distribution of AlOH \citep{Agundez2020}. However, the exact chemical pathways for Al-bearing species in star-forming regions remain uncertain, as most studies focus on the chemistry of AGB stars. This identification remains uncertain, as only one single transition of AlOH is covered in our observations. 
 We therefore present this feature as a possible AlOH transition identification, without claiming a detection of the molecule. Observations of multiple transitions at similar angular resolution are needed to confirm detection and clarify its relation to AlO and H$_2$O. This discussion is intended to motivate future observations designed to confirm the detection, particularly given the scarcity of observable AlOH transitions in the ALMA frequency range.

\bibliography{sample701}{}
\bibliographystyle{aasjournalv7}

\end{document}